\documentclass[structabstract]{aa}  
\usepackage{graphicx}
\usepackage{color}
\usepackage{xspace}
\def\rxj{RX J1131-1231\xspace}
\def\cha{Chandra\xspace}
\usepackage{txfonts}
\usepackage{natbib}
\def\gr{$\gamma$-ray}
\begin{document}

\title{Hard X-ray view of microlensing events in RX J1131-1231}

    \author{A. Neronov, D.Malyshev, R.Walter
          }

   \institute{ISDC, Astronomy Department, University of Geneva, Ch. d'Ecogia 16, 1290, Versoix, Switzerland \\
              }

\abstract
{RX J1131-1231 is a gravitationally lensed system which includes four images of a quasar lensed by an elliptical galaxy. The flux in the individual images is known to be affected by microlensing effect in the visible and X-ray bands. }
{We  study the multi-wavelength  properties of  RX J1131-1231 over a broad energy range, from optical to hard X-ray, during the periods of the microlensing caustic crossings. We aim to constrain the spatial extent of the X-ray emission region at different energies. }
{We combine the data of the source monitoring in the visible band with the X-ray data of the Burst Alert Telescope (BAT) on board of SWIFT satellite and Chandra X-ray observatory.    }
{Inspecting the broad band spectrum and lightcurves of the  source we identify several  microlensing caustic crossing events, and study the details of variability of the source during these  events. The caustic crossings of image A  on MJD 55150 and 55500 produce strong variations of the overall X-ray flux from the source.
In the soft X-ray band,  the caustic crossing events are characterised by the exponential growth / decay of the flux ratio between the source images with the exponentiating time scale $T_d\simeq 60$~d, which corresponds to the caustic  transit time  of the source of the size $\sim 10^{14}$~cm.  Variability induced by the microlensing dominates the overall variability of the sum of the fluxes of four images of the source which should also be detectable in the hard X-ray lightcurve, if the hard X-ray source is comparably compact.  Non-detection of the caustic crossing events in the SWIFT/BAT lightcurve of the source indicates that geometry of the region of emission of X-rays with energies above 15~keV is different from that of the lower energy X-rays, with the hard X-ray emission  an order of magnitude larger in size. }
{}

\maketitle

\section{Introduction}

Gravitational microlensing \citep[see e.g.][]{chang79,kayser86,schmidt10} is a powerful tool for the study of small, micro-arcsecond scale structures in remote quasars which are otherwise unresolvable to telescopes. It produces incoherent variability of the flux of individual images of the lensed quasar. The microlensing variability signal  is superimposed onto coherent variability of the images arising due to the intrinsic flux variations of the lensed source. Study of the microlensing effects in the visible \citep{eigenbrod08,dai10} , X-ray \citep{dai10,chartas12,macleod15} and \gr\ \citep{neronov15,vovk15} bands provides measurements of the sizes of accretion disks, hot X-ray coronae and \gr\ emitting jet launching regions near the supermassive black holes in quasars and potentially opens a way for testing relativistic gravity theory in the strong field regime \citep{neronov15a} in a way complementary to that of the study of gravitational waves from the compact object mergers \citep{gw,gw_test}.

\rxj is a quasar at the redshift $z=0.658$ lensed by a foreground galaxy at $z=0.295$ \citep{sluse03}. Strong gravitational lensing produces four images of the quasar, separated by $\simeq 1'' - 2''$ (see Fig~\ref{fig:image}) and referred hereafter as sources A--D. In the visible band, an image of the Einstein ring is also observed \citep{sluse03}. Both visible and X-ray fluxes of the four sources are variable, which makes the measurement of gravitational time delay possible. The flux from the images A,B and C arrives with nearly zero time delay, while the flux of the image D is delayed by 91~days~\citep{tewes13}. 
Precision of the time delay measurements (which are important in the context of the measurement of the Hubble parameter $H_0$ \citep{Suyu12}) is limited by the presence of microlensing effects which produce uncorrelated flux variation in different images \citep{tewes12a}. 

The X-ray spectrum of the source is composed of three components: the hard powerlaw, the soft excess and a  relativistically broadened Fe K$\alpha$ line, with close to the maximal possible angular velocity \citep{reis14}.  The X-ray flux of the four images is affected by the microlensing \citep{dai10,chartas12}. Statistical analysis of the microlensing variability, based on the method of \citet{kochanek04} leads to a constraint on the size of the X-ray source, $R_X\lesssim 10^{14}$~cm on the size of the source. This size is about 10-20 times the gravitational radius of  black hole of the mass $M_{BH}\sim 10^8M_\odot$ (which is an order-of-magnitude mass estimate for \rxj, see \cite{bh_mass_estimate}). 

Evidence for the energy dependence of the microlensing effect was reported by \citet{chartas12} based on the analysis of a microlensing event in the image D. A direct analysis of the energy dependence of the microlensing using the image D is complicated by the large time delay of the flux from the image. In the absence of X-ray observations done  with exactly 91~d time delay precludes the possibility to compare the spectrum of the image D, affected by the microlensing, with the simultaneous measurement of the intrinsic source spectrum spectra observed in other images of the source. In principle, difference between the spectrum of the source D and the spectra of other images measured in one and the same exposure could be due to the fact that the intrinsic source spectrum changes on the time scale of 91~d rather then due to the energy dependence of the microlensing effect. 

To test the hypothesis of the energy dependence of the X-ray microlensing effect one could rather compare the spectra of sources A, B and C, which arrive with nearly zero time delay \citep{tewes13}. This makes the simultaneous measurements of the intrinsic and microlensing-affected spectra straightforward.  Another possibility to test the energy dependence of the microlensing effect is to extend the energy range of the measurements. 

In what follows we use both approaches to verify the energy dependence of the microlensing effect. We identify the caustic crossing events of the image A, using the X-ray and visible band data, and study the caustic crossing induced variations of the source spectrum in X-rays. We complement the X-ray observations by \cha with the SWIFT/BAT observations, to study the influence of the microlensing effect on the hard X-ray flux.

\section{Chandra data analysis}
\label{sec:chandra}

The soft X-ray data used in the following analysis was obtained with the Advanced CCD Imaging Spectrometer (ACIS, see e.g.~\citet{garmire03}) on board of \cha X-ray Observatory. 39 observations of \rxj were taken between 2004-04-12 and 2014-07-12.

\begin{figure}
\includegraphics[width=\linewidth]{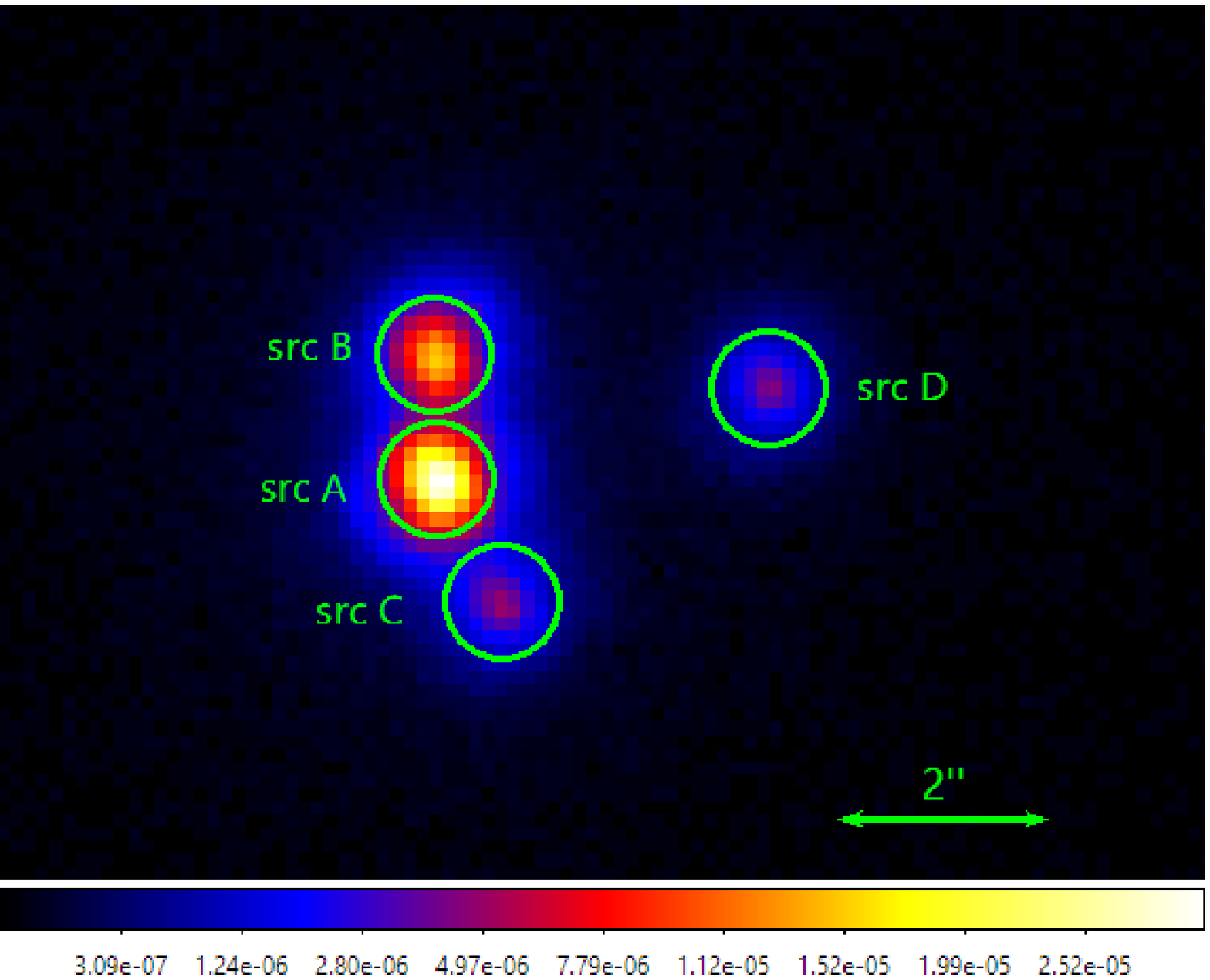}
\includegraphics[width=0.49\linewidth]{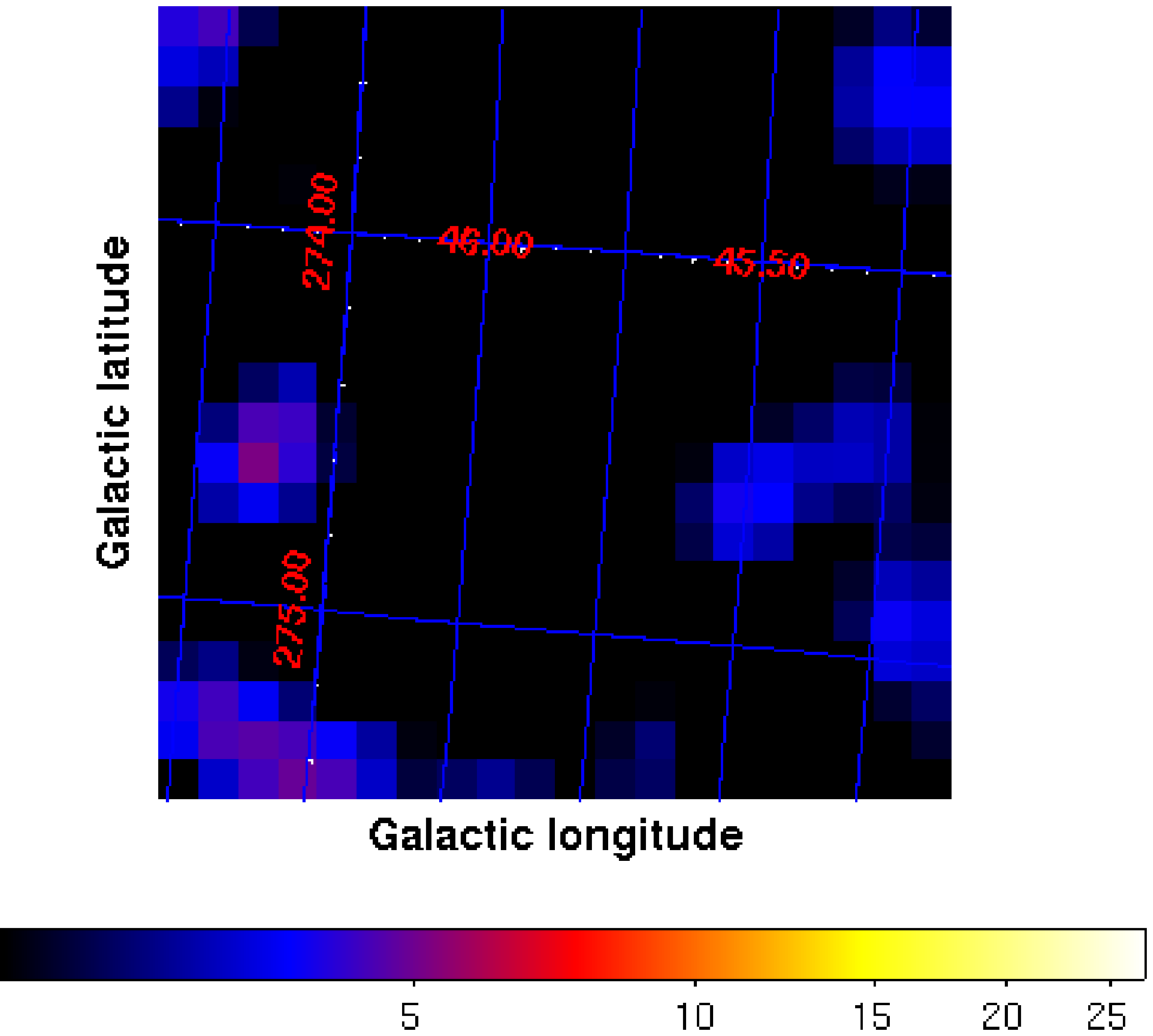}
\includegraphics[width=0.49\linewidth]{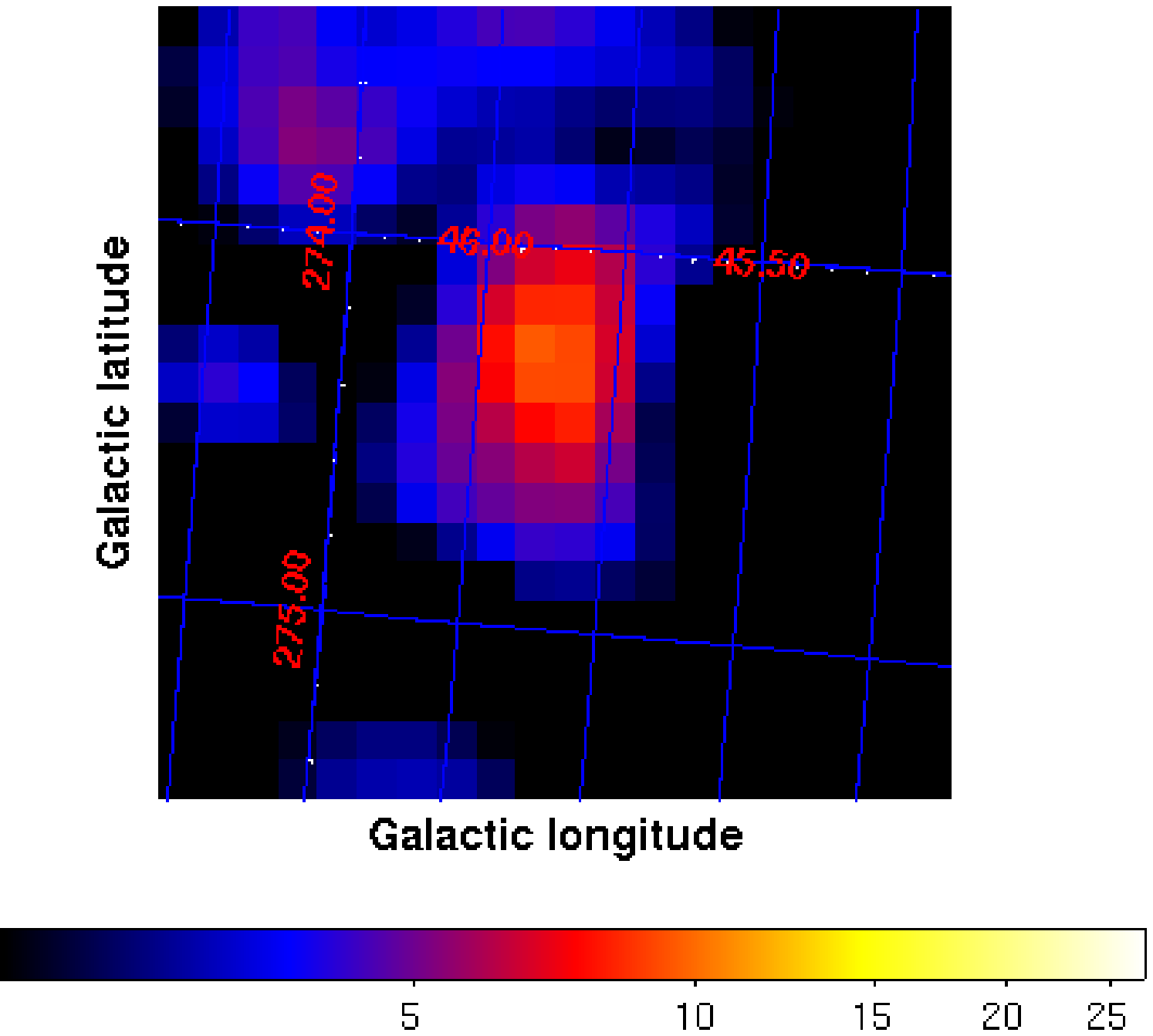}
\caption{Top:  staked image of \cha observations of \rxj in 0.6-9.0~keV band. The positions of 4 lensed images of \rxj are shown with green circles. The sizes of the circles correspond to the sizes of the regions used for \cha spectral analysis. Bottom: SWIFT/BAT images of \rxj for the time periods before 2007 and after 2008.} 
\label{fig:image}
\end{figure}

We perform the analysis using the standard CIAO v.~4.7 software provided by the \cha X-ray Center\footnote{http://cxc.harvard.edu/ciao/}, processing the data in the standard way, described in the data analysis threads. To improve the spatial resolution for the imaging analysis we use the sub-pixel resolution technique~\citep{tsunemi01}. The count map of \rxj obtained with this technique is shown in Fig~\ref{fig:image}. The circles (of 0.54'' radius) show the regions used for sources A--D spectra extraction, while the region used for the background estimation (3.2'' radius) is not shown in the figure. 

For the images B and C which are located at distances $\sim 1.3''$ from source~A one can expect the contribution from the tails of the PSF. In order to check the significance of this contribution for B and C sources we considered two different background selection regions. As an alternative to the 3.2'' radius background region we considered the composite region which consists of two circles 0.54'' radius each and located at distances 1.3'' from source A on the line connecting image A and image D.
We verified, that the spectral results are identical for both background selections. 

Following~\citet{reis14} we correct for the pile-up the spectrum of source A during the flaring state (MJD~55163 -- 55512). Pile-up is an instrumental effect that occurs when two or more photons strike the same or neighbouring CCD pixels within the integration time frame. As a result of the pileup the read-out electronics registers only one photon with the energy equal to the sum of energies of all involved photons. Thus the pileup leads to the apparent hardening of the slope of the source's spectrum. 
Performing the spectral analysis of the pile-uped observations we invoke the \texttt{pileup} XSPEC model~\citep{davis01}, combining it with the model of the emission from \rxj as described below. 

Fig.~\ref{fig:lightcurve} (middle panel) shows the 0.6--9.0~keV flux of image A and the sum of the fluxes of images  B and C. Bottom panel of the figure shows the sum of fluxes of all the images A-D as a function of time.  Each point on the lightcurve corresponds to one \cha observation of this source. In order to estimate the flux, during each observation we considered the simple absorbed powerlaw model in 0.5-10~keV energy band. For the observations taken during the flaring state we additionally include into the model the \texttt{pileup} component. In order to estimate the flux for these observations after the fitting we remove this component from the model\footnote{See e.g. https://heasarc.gsfc.nasa.gov/xanadu/xspec/manual/XSmodelPileup.html}. 
The $1\sigma$ uncertainty on the measured flux was estimated as the standard deviation of the mean flux values obtained in $\sim 500$ realizations of the best-fit source A model. 
We find, that the pileup-corrected fluxes are significantly higher (up to a factor of $\sim 2$ for the brightest observation 11540) than one obtained without pileup assumption.

In all other cases we derive the flux in the corresponding energy range convolving the considered model with \texttt{cflux} component and performing the fit. The shown uncertainty on the flux stands for $1\sigma$ confidence level and corresponds to the change of the best-fit $\chi^2$ value by 3.5~\citep[see e.g.][]{lampton76}


\section{BAT data analysis}
The burst alert coded mask telescope (BAT, \citet{krimm13}) on board of the Swift satellite has a field of view of 1.4 sr for a point spread function of 17~arcmin. Swift performs typically 50-100 short pointings in different directions every day and BAT thus monitor the complete sky at hard X-rays every few hours.
For very bright sources, lightcurves can be obtained in different energy bands with a resolution of 1000~sec for almost a complete decade.

The Swift/BAT reduction pipeline is described in~\citet{tueller10} and~\citet{baumgartner13}. Our pipeline is based on the BAT analysis software HEASOFT v.6.13. A first analysis was performed with the task \texttt{batsurvey} to create sky images in the 8 standard energy bands using an input catalogue of 86 bright sources (that have the potential to be detected in single pointings) for image cleaning. Background images were derived removing all detected excesses with the task \texttt{batclean} and weighted averaged on a daily basis. The variability of the background was then smoothed pixel-by-pixel using a polynomial model with an order equal to the number of months in the data set. The BAT image analysis was then run again using these averaged background maps. The image data were stored in a database organised by sky pixel (using a fixed pixel grid) by properly projecting the images on the sky pixels, preserving fluxes. This database can be used to build images, spectra or lightcurves for any sky position and time sampling.

The result of our processing was compared to the standard results presented by the Swift team (lightcurves and spectra of bright sources from the Swift/BAT 70-months survey catalogue\footnote{http://swift.gsfc.nasa.gov/results/bs70mon/}) and a very good agreement was found. 

The Swift/BAT lightcurve of RX J1131-1231 was built in the energy band 14-75 keV. For each time bin a weighted mosaic of the selected data is first produced and the source flux is extracted assuming fixed source position and shape of the point spread function. The source signal to noise varies regularly because of its position in the BAT field of view and distance to the Sun. The lightcurve was extracted from mosaic images built accumulating successive periods of 150 days  to allow for source detection. The averaged spectrum of the source was also extracted from mosaic images. 

\begin{figure}
\includegraphics[width=\linewidth]{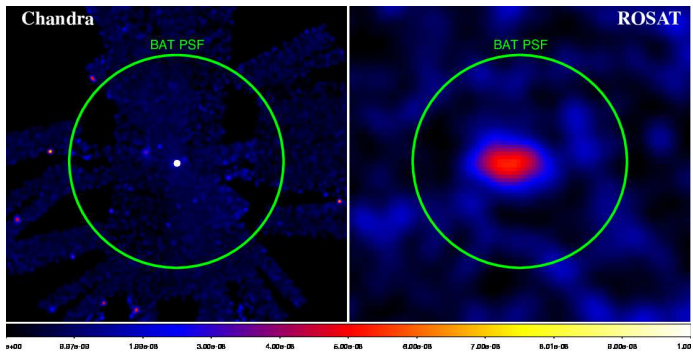}
\caption{Mosaic image of \cha observations of the \rxj field in the energy band 2-7~keV (left) and ROSAT all-sky survey image in the 0.1-2.4~keV band (right), compared to the BAT PSF.} 
\label{fig:image_wide}
\end{figure}

The BAT images of the source are shown in the bottom panels of Fig. \ref{fig:image} for two time periods: before 2007 and after 2008. BAT PSF is much wider than the angular separation of the individual images of \rxj. This wider PSF might also lead to the source confusion: a nearby unrelated source could contribute to the BAT signal. We have verified that there are no bright X-ray sources other than \rxj within the BAT PSF extent by building a mosaic of \cha observations and also checking the ROSAT all sky survey image (see Fig. \ref{fig:image_wide}).  

\section{Timing analysis: microlensing events of image A}

During the decade long period considered in our analysis the strongest image of the source was image A, see Fig. \ref{fig:image}. A natural assumption is that the contribution of the source A also dominates the total hard X-ray flux of the source. Taking into account that BAT is not able to resolve individual images of the source, the only possibility to study the influence of the microlensing on the hard X-ray flux is to identify the microlensing events of the strongest image A and study the spectral variability of the source during these events. 

The top and middle panels of Fig. \ref{fig:lightcurve} show the lightcurves of image A and of the sum of images B and C in the visible and in the X-ray band. The overall variability is composed of the intrinsic variability of the source and variability related to the microlensing. 

Moderate scale of variability observed in  images B and C indicates that the intrinsic flux variability of the source was by no more than $\pm 50$\% in X-rays. A consequence of this conclusion is that strong variability of the flux of the image A should be attributed to the microlensing. The amplitude of the microlensing variability in image A has reached a maximum on about MJD 55153, in \cha observation ID 11540, when the flux grew by a factor more than 5 compared to the previous observation. A natural interpretation of such a strong transient increase of the microlensing effect is due to the crossing of a microlensing  caustic in front of the image A. The caustic crossing episode around this date could also be readily identified in the visible band lightcurves \citep{tewes13}. The ratio of the flux of the image A to the average flux of the images B and C shown by the black points in the top panel of Fig. \ref{fig:lightcurve} shows a clear maximum in the period around MJD 55100. Although the maximum is reached during a gap in the observational campaign, the position of the maximum could be localised by fitting the growing and decaying part of the flux ratio curve, as it is shown in Fig. \ref{fig:lightcurve}, top panel. 

An additional maximum of the flux ratio of the image A to the average flux of the images B and C is found in the X-ray band on MJD 55511, during the \cha observation ID 11544. The visible band counterpart of this maximum could also be identified in the flux ratio curve shown in the top panel of Fig. \ref{fig:lightcurve}.  Finally, still one more maximum of the ratio of the flux of the image A to the average flux of B and C is observed on MJD 54100 during \cha observation ID 7785.

Temporal evolution of the X-ray  flux ratio of image A to the average flux of images B and C exhibits fast variations during the microlensing episodes on MJD 55100 and MJD 55511. During this episodes the ratio grows / decays exponentially with the exponentiating time scale $T_s\simeq 60\pm 30$~d (see  model  black curve in the middle panel of Fig. \ref{fig:lightcurve}). Measurement of the time scale of the variability of the flux ratio at the moment of the caustic crossing provides an estimate of the size of the X-ray source
\begin{equation}
\label{eq:size_soft}
R_s\sim T_sv_{proj,\bot}\simeq (1.5\pm 0.8)\times 10^{14}\left[\frac{v_{proj}}{300\mbox{ km/s}}\right]\mbox{ cm} 
\label{eq:soft}
\end{equation}
where $v_{proj,\bot}$ is  the projected velocity of the source, the caustic and the observer in the direction normal to the direction of the caustic.
The velocity scale $v_{proj,\bot}\sim 300$~km/s is characteristic for the motion of stars in the Milky Way mass scale galaxies. It determines the motion of the Solar system (and of the observer on the Earth) and the motion of stars in the lensing galaxy (and of the microlensing caustics pattern).

The scale of flux magnification during the microlensing caustic crossing episodes is determined by the source size $R_s$ relative to the size of the Einstein ring of the microlensing masses, $R_E$. It scales as $\sqrt{R_E/R_s}$. An order-of-magnitude larger size of the visible band source leads to a smaller scale variations of the flux ratios during the microlensing events, as one could see from the top panel of Fig. \ref{fig:lightcurve}. The model of the evolution of the flux ratio of the image A to the average flux of images B and C shown in this panel consists of two components. The dashed curve shows the model component obtained by convolving the soft X-ray flux ratio model from the middle panel of Fig. \ref{fig:lightcurve}  with a Gaussian of the width $T_V=1200$~d (corresponding to the extent of the visible band source \citep{dai10}. The solid curve shows the sum of this slowly variable flux ratio model with the fast variable component which just follows the soft X-ray flux ratio variability. Physically this fast variable component corresponds to the emission from the innermost part of the accretion disk (of the size comparable to the size of the X-ray source) which has higher temperature.

\section{Timing analysis: SWIFT/BAT lightcurve}

\begin{figure}
\includegraphics[width=\linewidth]{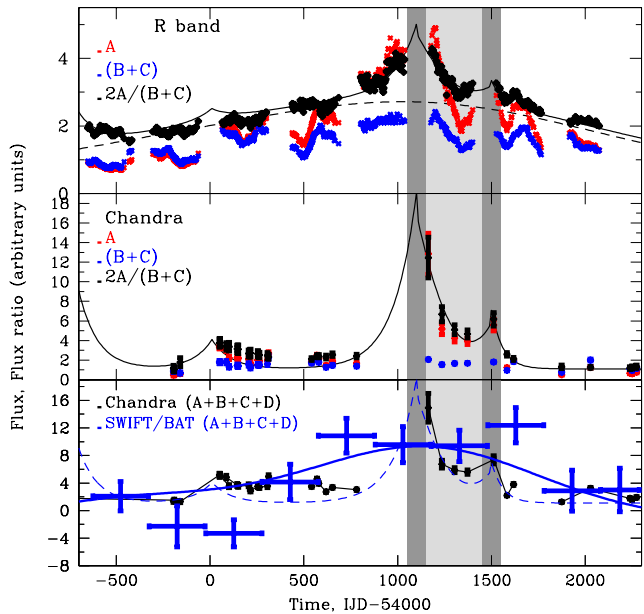}
\caption{Visible (top), X-ray (middle) and hard X-ray (bottom) lightcurves of the source. Blue data points in the bottom panel show the BAT lightcurve. Solid curve show the best fit model calculated assuming that the hard X-ray source is the same size as the visible band source. Dashed line shows the best fit model assuming that the hard X-ray source is the same as the soft X-ray source.  Solid curve in the middle panel shows a model for the flux ratio $M=2F_{A}/(F_B+F_C)$ described in the text. In the top panel the dashed curve shows the model curve $M$ convolved with a Gaussian of the width $\Delta T=1200$~d. Solid curve shows the microlensing flux ratio model for the sum of the compact and extended source. Grey shaded ranges highlight the main microlensing episodes, with the darker shading corresponding to the caustic crossings.} 
\label{fig:lightcurve}
\end{figure}

A conventional assumption for the origin of the X-ray emission in radio-quiet quasars like \rxj is that the hard powerlaw component of the flux is produced by the ``hot corona'' (i.e. a part of the accretion flow with temperature much higher than that of the optically thick accretion disk). In principle, the geometry of the soft and hard X-ray emitting part of the corona is assumed to be identical. In such situation,  the influence of the microlensing on the hard X-ray flux should be the same as on the soft X-ray flux. One expects to detect a large enhancement of the hard X-ray flux of the image A during the strongest microlensing episode on MJD 55150. 

The angular resolution of SWIFT/BAT is not sufficient to resolve separate the flux of the image A from the fluxes of other images B-D. BAT could measure only the overall source flux, which is the sum of the fluxes of images A-D.  The strongest microlensing episode around MJD 55150 has also  produced  the overall flux enhancement of the flux of the sum of the four source images (A+B+C+D), as one could see from the bottom panel of Fig. \ref{fig:lightcurve}. The enhancement was by a factor of $\simeq 4$ compared to the source flux level prior to the microlensing episode. If the size of the hard X-ray source is the same as of the soft X-ray source,  a strong flux enhancement should also be observed in the hard X-ray band by SWIFT/BAT.

Fig. \ref{fig:lightcurve} shows lightcurve of the source in the energy range 15-75~keV.  One could see that no obvious flux enhancement during the microlensing episode on MJD 55150 is observed. A model of the BAT lightcurve provided by the soft X-ray flux evolution (dashed blue curve in the bottom panel of Fig. \ref{fig:lightcurve}) is inconsistent with the BAT data  (the $\chi^2$ of the fit of the BAT lightcurve with the soft X-ray model lightcurve is 80 for 8 degrees of freedom). 

The SWIFT/BAT lightcurve is also inconsistent with constant flux hypothesis, the fit has $\chi^2=47$ for 9 degrees of freedom.  A better  fit ($\chi^2$ of the fit 22.7 for 8 degrees of freedom)  is provided by the model obtained from the soft X-ray band lightcurve model via convolution with a Gaussian of the width $T_{h}=600_{-200}^{+400}$~d. Measurement of the microlensing time scale  in the hard X-ray band, obtained in this way, provides an estimate of the hard X-ray source size
\begin{equation}
R_h\sim T_{h}v_{proj}\simeq 1.5_{-0.5}^{+1.0}\times 10^{15}\left[\frac{v_{proj}}{300\mbox{ km/s}}\right]\mbox{ cm.} 
\label{eq:hard}
\end{equation}
This is an order-of-magnitude larger than the size of the soft X-ray source and is comparable to the size of the visible band source. 

An alternative origin of variability of the source hard X-ray flux could be the intrinsic variability of the source (still, the microlensing variability should be superimposed on the intrinsic flux variations). The intrinsic variability in hard X-rays could be traced by the soft X-ray spectral variability of the source components. For example, softening of the source spectrum during the period before 2008 could be responsible for the low hard X-ray flux during this period. We have verified that the intrinsic source spectrum does experience variations between \cha observations, but there is no trend which would explain the variability of the hard X-ray flux. In particular, the intrinsic spectrum (with the photon index varying in the 1.7-1.8 range) was not softer before 2008. Thus,  most of the long-term variability of the hard X-ray flux of the source should still be ascribed to the influence of the microlensing.

\section{X-ray spectral analysis}

Comparison of the strength of the microlensing effect in soft and hard X-ray bands reveals a difference in the sizes of the soft and hard X-ray sources. There are different possible source geometries which would accommodate such a difference. 

First, the source size could change gradually with energy with lower energy photons emitted closer to the black hole. More efficient cooling of denser material in the innermost part of the hot corona  could lead to the lower average energies of electrons closer to the black hole and, as a result, lower energy of X-rays emitted by these electrons.  Alternatively,  soft and hard X-ray fluxes could be dominated by different emission components, e.g. by the Bremsstrahlung from the hot corona and Compton reflection from the accretion disk, which occurs on different distance scale. These two possibilities could be distinguished through the measurement of the details of the change of the microlensing variability pattern as a function of energy and emission component. 

\begin{figure}
\includegraphics[width=\linewidth]{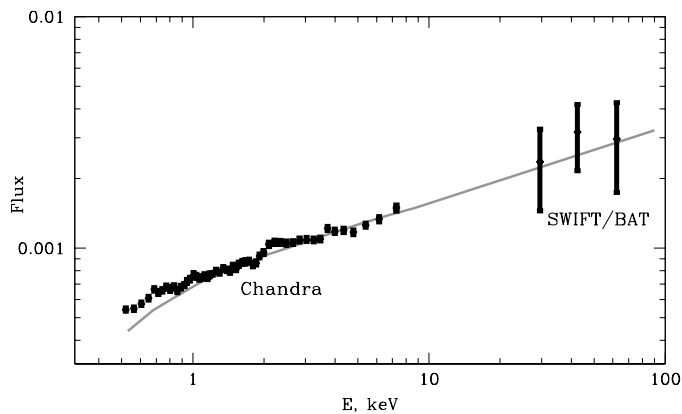}
\caption{Time averaged X-ray spectrum of \rxj.}
\label{fig:spectrum}
\end{figure}

 Fig. \ref{fig:spectrum} shows the time-averaged combined spectrum of the images A-D for the entire observation period excluding the time interval between MJD 55150 and MJD 555200, where the source flux is affected by the pileup effect in the image A. The spectrum contains  different components: the broadband powerlaw originating from the hot corona, the soft excess, the broadened  fluorescence iron line from the innermost part of the accretion disk \citep{done12,reis14}. The model curve in the figure shows a powerlaw component of the source flux stretching from soft to hard X-ray band. One could see that the BAT flux measurements lie at the extrapolation of the soft X-ray spectrum. The best fit powerlaw model has the  slope $\Gamma=1.70\pm0.02$. An excess over the powerlaw in the 2-4 keV energy range could be attributed to the presence of the narrow / broad iron line, which we model as a sum of two gaussians: one narrow at the nominal energy of the Fe K$\alpha$ line at   $6.4/(1+z)\simeq 3.8$~keV  and a broad line at the energy $2.31\pm 0.06$~keV with the width $0.25\pm 0.08$~keV. The spectrum below 1~keV shows an excess over the powerlaw absorbed by the Galactic hydrogen column density $N_H=3.6\times 10^{20}$~cm$^2$ \citep{nh}. This could be interpreted as a ``soft excess''. Following \citet{reis14} we model the soft excess with the {\tt diskbb} model with the inner temperature $T=0.21\pm 0.02$~keV. The reduced $\chi^2$ of the fit is $\chi^2=1.18$ for 419 degrees of freedom. 
 
 In the hard X-ray spectrum measured by BAT (see Fig. \ref{fig:spectrum}) no signature of the high-energy cut-off is detected.  A 2$\sigma$ lower bound on the cut-off energy is $E_{cut}\ge 50$~keV. The reflection component, which should appear as a "bump" in the source spectrum in the energy range about 30~keV, is not clearly visible, although its sizeable contribution is not excluded.

If the strength of the microlensing variability gradually decreases with the increase of photon energy, one would expect, that during the microlensing event the spectrum of the lensed source becomes softer. The simplest quantitative measure of the spectrum which can demonstrate this effect is the hardness ratio $H=F_{soft}/F_{hard}$, where for soft and hard energy bands we adopt 0.6-2~keV and 2-9~keV correspondingly. However, besides the microlensing, the hardening/softening of the observed spectrum can be explained by the intrinsic source's variability. In order to correct for this one can utilise the hardness ratio of non-lensed images (B and C) and during the microlensing event of e.g. image A to consider the normalized hardness ratio $\cal H_{A,B+C}=H_{A}/H_{B+C}$. It should be equal to 1 if the spectrum of the source A is proportional to the spectrum of B and C sources. The deviation of this quantity from 1 during the microlensing episode can not be explained by intrinsic variability of \rxj, since there is no time delay between A,B and C images. 

In fact, \citet{chartas12} have reported a significant ($1.3 \pm 0.1$) deviation  of $\cal H$ for the source D during the microlensing event at MJD~55200--55500. As it is mentioned in the introduction, a potential problem of analysis of \citet{chartas12} was in neglecting the large, 91~d, time delay of the flux coming from the image D for calculation of this quantity. 

\begin{figure}
\includegraphics[width=\linewidth]{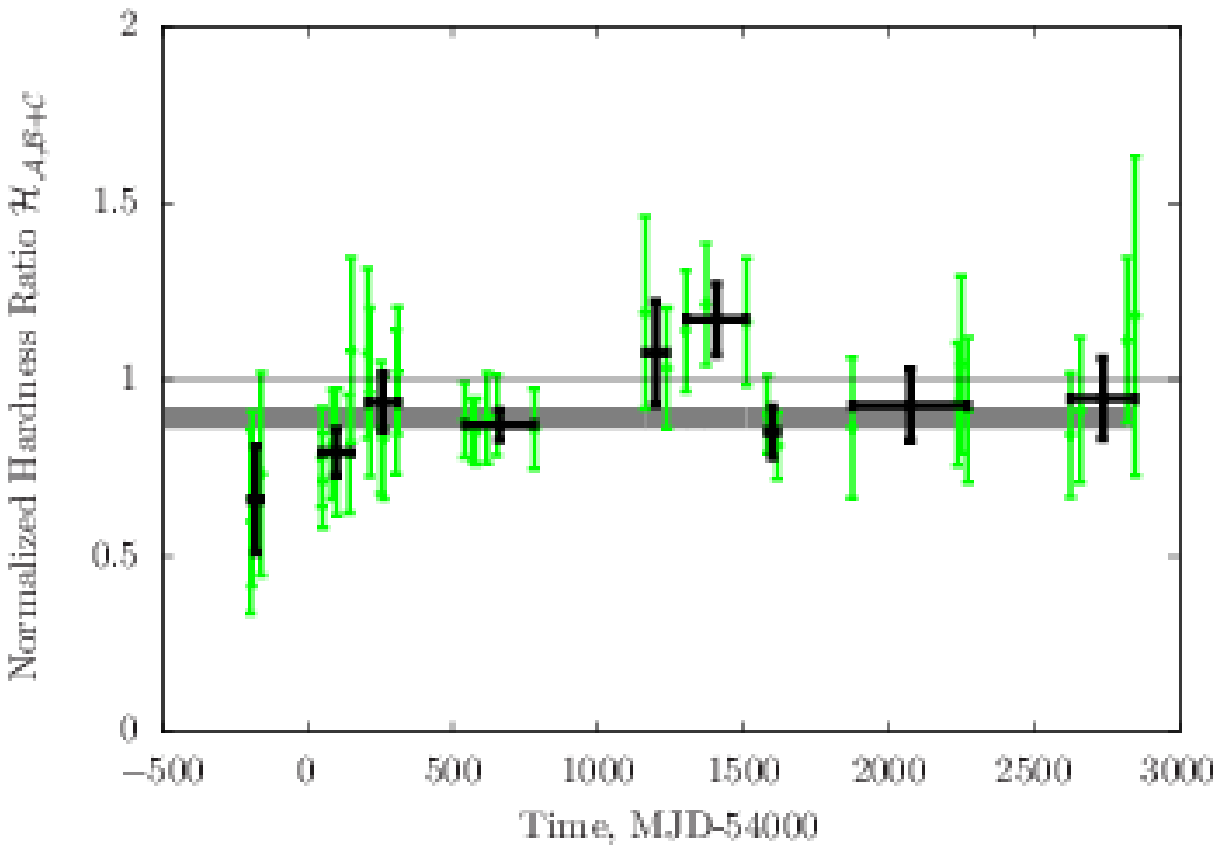}
\includegraphics[width=\linewidth]{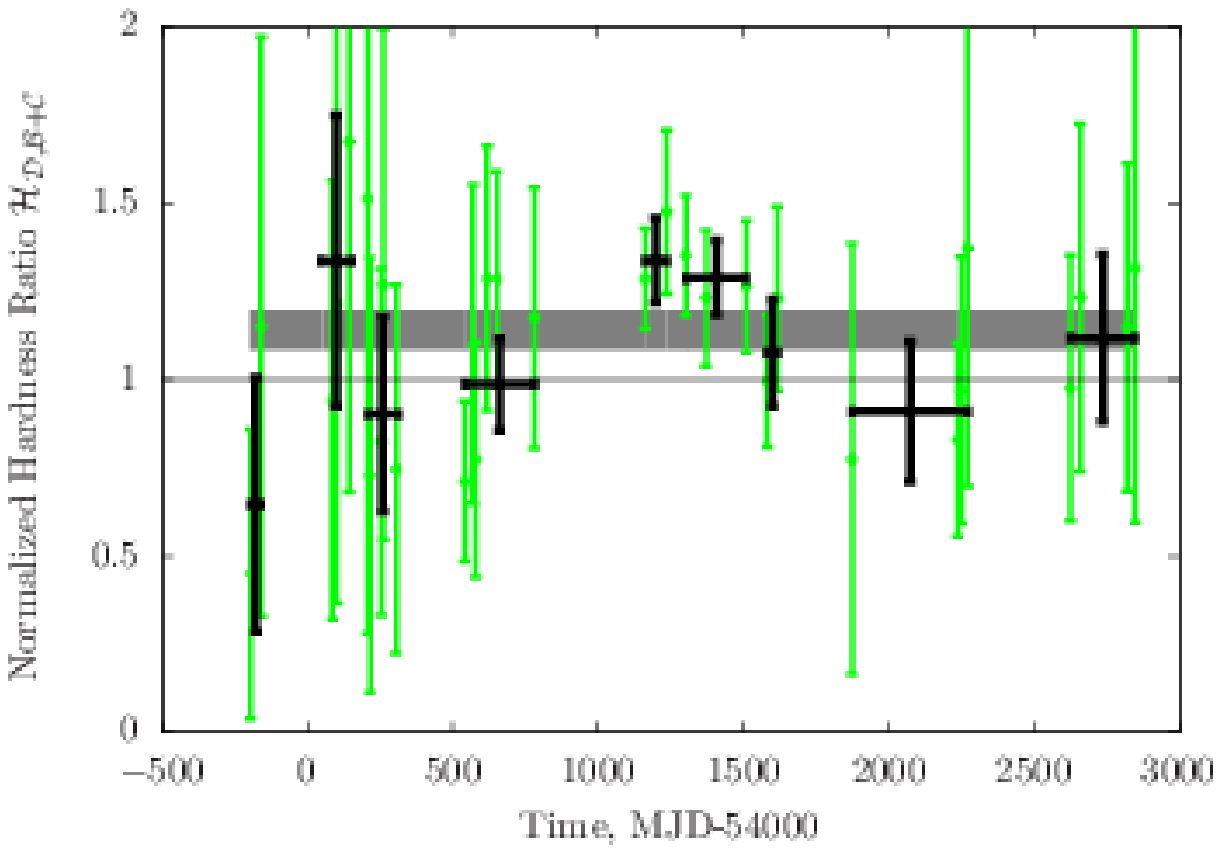}
\caption{The evolution of the normalized hardness ratios $\cal H_{A, B+C}$ (top panel) and $\cal H_{D, B+C}$ (bottom panel), see text for the details. The thick green points with the errorbars show the normalized hardness ratios for each individual \cha observation of \rxj, while the solid green -- for grouped in time observations. Shaded area indicates uncertainties of the best-fit of the normalized hardness ratios with the constant. }
\label{fig:hardness}
\end{figure}

In Fig.~\ref{fig:hardness} we show the normalized hardness ratios for images A ($\cal H_{A,B+C}$, top panel) and D ($\cal H_{D, B+C}$, bottom panel). The time delay of the flux from image D is taken into account in the calculation of ${\cal H}_D$ via linear interpolation of the time shifted flux measurements to the dates of observations of the fluxes of images B and C. Thin green points with errorbars show the normalized hardness ratios measured in each single \cha observation of \rxj while the thick black ones -- the results for grouped nearby in time observations.

If the microlensing effect would be energy-independent all over the \cha energy band, both $\cal H_{A,B+C}$ and $\cal H_{D, B+C}$ would always be equal to 1. Deviations from 1 observed in both normalized hardness ratios clearly show that the influence of the microlensing on the source flux is energy dependent already if \cha data alone are considered (thus confirming the result of \citet{chartas12} after correction for the time delay of the image D).

The quality of \cha data is not sufficient for a more detailed assessment of the energy dependence of the microlensing variability during the caustic crossing events. In general, both ${\cal H}_A$ and ${\cal H}_D$ are consistent with being constant in time (but not equal to one). The constant fits to ${\cal H}_A$ and ${\cal H}_D$ and their uncertainties are shown by the grey shaded areas in Fig. \ref{fig:hardness}. Although the overall deviation of  ${\cal H}_A$ and ${\cal H}_D$ from unity provide certain measure of the energy  dependence of the microlensing effect, this measure is difficult to characterise.  More straightforward would be interpretation of measurement of ${\cal H}_A$ and ${\cal H}_D$ in particular time intervals.

In the case of the image D, the most significant deviation of ${\cal H}_D$ from 1 occurs during the caustic crossing episode around the time MJD 55200 -- MJD 55500, considered by \citet{chartas12}.  The value ${\cal H}_D = 1.29 \pm 0.07$ observed during this episode corresponds to $\simeq 30\%$ decrease of the microlensing effect with the factor-of-two increase of the photon energy (the average energy of photons in the $0.6-2$~keV band is about 1~keV, most of the photons in the $2-9$~keV band have energies close to the lower energy boundary of the bin. Extrapolating this trend to higher energies (factor of two decrease of the variability amplitude for a ten-fold increase of photon energy) one could find that the microlensing should produce moderate flux variations in the SWIFT/BAT band, as observed. Changing the boundaries of the soft/hard energy bands for which we calculated the normalized harness ratio $\cal H_{D, B+C}$ to $0.5-1$~keV and $3-9$~keV we find a marginal confirmation of this statement. For the modified energy ranges, the corresponding value of $\cal H_D$ during the caustic crossing appeared to be $1.43\pm 0.09$.

The largest deviation of the value of ${\cal H}_A$ from unity is observed during the period around MJD 54000. \citet{dai10} have previously considered a possibility that during this period a caustic crossing of image B has occurred. This could explain the value ${\cal H}_{A}\simeq 0.7$  via stronger magnification of the flux of the image B in the $0.6-2$~keV band during this period, assuming the same energy dependence of the microlensing variability (a factor of 2 decrease per decade of energy) as derived for the image D. 

${\cal H}_A$ is formally consistent with one during the period of caustic crossing of image A on MJD 55100. However, the general evolutionary trend of ${\cal H}_A$  over the time interval preceding the caustic crossing suggests that this coincidence might be accidental. 

\section{Discussion}

A combination of visible, X-ray and hard X-ray data on long-term monitoring of \rxj\ provides a coherent view of the the source behaviour with respect to the microlensing. Monitoring of the source on the time scale of about 10 years reveals several events of rapid variability of the flux ratio of the three brightest images of the source, A, B and C. These fast variability episodes could be identified with the caustic crossing events. The caustic crossing events of the brightest image A produce significant variations of the overall flux of the source (the sum of the fluxes of A-D), so that these caustic crossing events could be identified also with a telescope which does no have angular resolution sufficient for separation of the four images. 

We have used this fact to characterise the effect of the microlensing on the hard X-ray flux of the source in the energy band 15-80~keV measured by SWIFT/BAT. We have found that the amplitude of the microlensing induced variability in this energy band is much lower than that in the 0.5-9~keV band. 

This indicates that the size of the hard X-ray source $R_h$ is much larger than that of the soft X-ray source, $R_s$, with the estimates given by Eqs.~(\ref{eq:soft}) and (\ref{eq:hard}). The soft X-ray source size estimate $R_s\sim 10^{14}$~cm found from the measurement of the exponentiating time scale of the flux during a particular caustic crossing episode agrees well with the estimate derived by \citet{dai10} based on statistical analysis of the microlensing induced variability in the X-ray band. 

Conclusion about the order-of-magnitude larger size of the hard X-ray source detected in the SWIFT/BAT band, $R_h\sim 10^{15}$~cm, is supported by the analysis of the energy dependence of the microlensing variability in the \cha energy band. \cha data indicate that the amplitude of the microlensing variability diminishes by a factor of 2 (and the source size increases by a factor of four) when the photon energy increases by an order of magnitude.

The quality of available \cha and SWIFT/BAT data is only marginally sufficient for the study of the details of the energy dependence of the microlensing effect (and hence of the energy dependence of geometry of the X-ray source). Higher quality measurements could be achieved via more detailed studies of the spectral evolution of the source images during the microlensing caustic crossing events. Such detailed studies could be done for the caustic crossings of the brightest image even with telescopes which could not resolve the individual images, such as XMM-Newton in soft X-ray and NuSTAR or Astro-H in the hard X-ray band. 

Studies of the spectral variability of the source during the caustic crossing episodes could provide valuable information on the largely uncertain  geometry of the hot X-ray emitting corona which dominates the X-ray flux of the radio quiet quasars and Seyfert galaxies. It could also help to clarify the origin of the soft excess component of the X-ray flux, which is also largely uncertain \citep{gierlinski04,fabian05,rozanska15}.


\end{document}